\documentclass{article}
\usepackage{rotating}
\usepackage{graphicx}
\graphicspath{ {./} }
\usepackage{hyperref}
\usepackage[
backend=biber,
style=alphabetic,
]{biblatex}
\title{Crystal structure discrimination based on a single atom speed dynamics}
\begin{document}
\author{Rafa\l \space Abram \and Dariusz Chrobak}
\maketitle

\begin{abstract}
Atom arrangement plays a critical role in determining material properties. It is, therefore, essential for materials science and engineering to identify and characterize distinct atom configurations. Currently, crystal structures can be determined either by its static properties or by quantifying its structural evolution. Here we show how to classify an atom into phase solely by its speed dynamics. We model silicon crystals at different phase transition points and use a single atom speed trajectory to demonstrate that crystal-structure-independent Maxwell distribution of speed is generated by crystal-structure-dependent atom dynamics. As the classification accuracy of the method increases with trajectory length, we show that subtle difference in local atomic structures can be identified using sufficiently long trajectories. Thanks to symbolization of atom dynamics, the method is computationally efficient and suitable for an analysis of large datasets on the fly.
\end{abstract}

\section{Introduction}
An accurate discrimination of local atomic structures of materials is necessary to understand structure-function relationship. Contemporary experimental methods have enabled impressively precise structural material characterization at real time \cite{Alcorn2023} and have therefore motivated a dynamic progress in an analy\-sis of material datasets coming from both experiments and simulations. In particular, growing interest in probing and quantifying physical processes calls for methods suitable to study temporal structures of evolving systems and its relations to other system's properties, eg. spatial structure. Methods of identifying distinct atom arrangement based on dynamic (and static) aspects of the systems have been recently reviewed \cite{Tanaka2019, Dijkstra2021, Zeni2021, Leitherer2021, Xie2019, Allera2024a}, including ordered and disordered materials. Additionally, an interesting concept of Chaotic Crystallography was proposed by Varn et al. \cite{Varn2015, Varn:ib5012} and applied to classify material structures using, among others, the stationary process describing a material that plays an analogous role to the lattice in classical crystallography. Similarly, Ryabov et al. used computational mechanics \cite{Crutchfield2012} to demonstrate that symbolic dynamics of a single particle velocity can be utilized to quantify transport properties in high-dimensional molecular systems, assuming systems's ergodicity \cite{Ryabov2011}. Generic methods of identifying distinct particle's environment based on a single particle track have been attracting researchers across many disciplines and have therefore been studied as part of Anomalous Diffusion challenge \cite{Muoz-Gil2021} The results of the open contest also showed that the temporal structure of a single particle track has an impressive power in identifying both an underlying diffusion model, its parameters and change points, even for non-ergodic processes and an extremely short trajectory's length

The aim of this research is to investigate the extent to which a single atom dynamics encodes its local neighborhood in crystalline materials at equilibrium. As an example, we analyze silicon atom trajectories, as its phases are broadly used \cite{Gerbig2016, Wong2019}, bc8 and r8 structures thereof are difficult to distinguish \cite{10.1103/PhysRevB.101.245203} and r8 has two non-equivalent Wyckoff positions that can be, in principle, further discriminated. As a local atom arrangement depends on both temperature and pressure of the sample, we setup our computer experiments at phase transition points making crystal structure the only difference among coexisting atom arrangements. Additionally, we symbolize trajectories to efficiently process the long ones and use permutation entropy \cite{10.1103/PhysRevLett.88.174102} to discriminate trajectories by quantifying their temporal structures. As the methods based on permutation patterns have recently been extensively studied, the interested readers will find several reviews and open-source software as well \cite{Pessa2022, Zanin2021, DatserisParlitz2022, Pessa2021, Datseris2018}.

The reminder of this paper is organized as follows: In Section \ref{methods}, we calculate phase transition points, describe the protocol of generating atom trajectories from molecular dynamics simulations and shortly introduce permutation entropy method. Subsequently, in Section \ref{results}, an entropy-based crystal structure discrimination scheme is presented and followed by the demonstration of how the accuracy of the method increases with trajectory length enabling identification of subtle differences in local atom neighborhood. Finally, in Section \ref{discussion}, we discuss applications of the method, its current limitations and outline further development. All the data, input scripts and codes for replicating figures presented here can be found on github: \url{https://github.com/rmabram/crystal-structure-classification}.

\section{Methods}\label{methods}
In this work we used two computational models of bulk silicon developed by Kumagai et al. \cite{Kumagai2007} and Y. Zuo et al. \cite{Thompson2015}. The former one is used to model diamond, bct5, btin and bc8 crystal structures, as it is computationally efficient and the simulated phases are stable at its approximate phase transition points. The latter one, in turn, we use to simulate bct8 and r8 structures, as both phases are similar and require more accurate model that correctly yields two Wyckoff positions of the r8 phase. All silicon samples considered here are modeled at 300K and different pressures: diamond and bct5 structures at 10 GPa, bct5 and btin at 16.5 GPa, btin and bc8 at 7GPa, and both bc8 and r8 at 4.5 GPa. The values of the pressure were calculated in Supplementary materials, Section I, and correspond to experimental phase transition points determined by Wong et al. \cite{Wong2019}. An isothermal-isobaric ensemble and Nose-Hoover dynamics \cite{10.1103/PhysRevB.69.134103} are used with the time step of 2 fs, pressure damping parameter of 500 fs and temperature damping parameter of 30 fs. All simulations were preceded by sample's energy minimization using conjugate gradient algorithm and each sample was subsequently equilibrated for 200 ps. Finally, 20 ps long trajectories of 56 randomly selected atoms were recorded for diamond, bct5, btin and bc8 samples, whereas 1 ns long trajectories of atoms from the piece of r8 and bc8 phases were used to investigate and visualize Wyckoff positions. LAMMPS \cite{Thompson2022} input scripts defining our molecular dynamics simulations can be found either on GitHub (url{https://github.com/rmabram/crystal-structure-classification}) or in Supplementary materials, Section II. Atom trajectories were extracted from the LAMMPS output files using an open source MDAnalysis Python package \cite{Michaud-Agrawal2011}. Crystal structures were visualised using OVITO software \cite{ovito}. 

A symbolic speed trajectory that is used throughout this work is defined as a sequence of unique permutation patterns (or ordinal patterns), each of which constructed as follows: given a set of four consecutive values of speed, $s_t=\{v_t, v_{t+1}, v_{t+2}, v_{t+3}\}$, sort it and replace every value of speed with its position in the initial set, $s_t$. (As there are only four positions, the maximum number of permutation patterns equals to $4!=24$.) Then, the symbolic trajectory is generated by building a sequence of the ordinal patterns of every four consecutive speed values from the trajectory. It is performed using an \texttt{ordinal\_sequence} method from ordpy Python package developed by Pessa et al. \cite{Pessa2021}. A set of four consecutive values of speeds is a minimum one that provides clear and satisfactory results for all atom speed trajectories considered here. Finally, using \texttt{permutation\_entropy} method from the ordpy package we calculate permutation entropies, i.e. Shannon entropy \cite{Shannon1948} of an ordinal patterns probability distribution, in order to discriminate atom speed dynamics.

\section{Results}\label{results}
\begin{sidewaysfigure}[h] 
\centering
\includegraphics[width=\columnwidth]{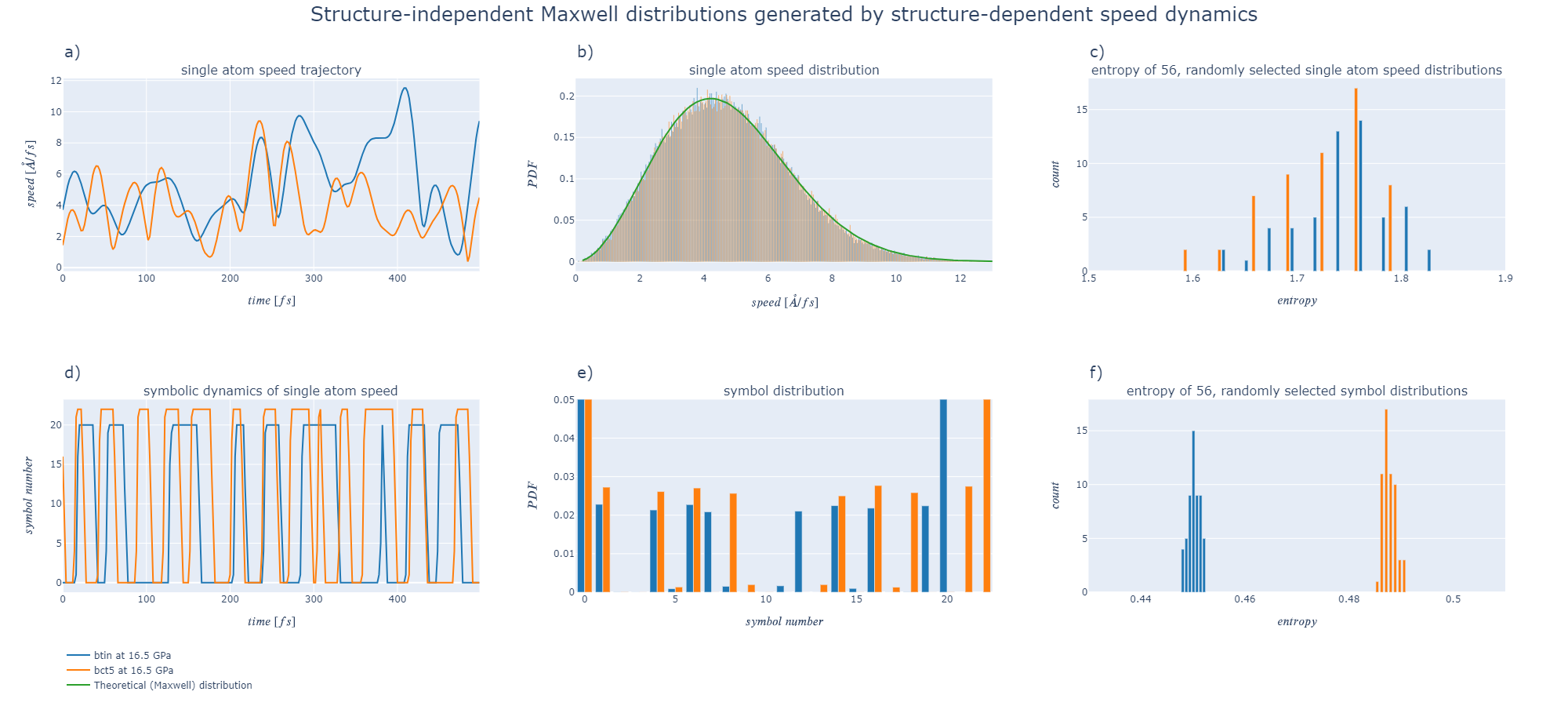}
	\caption{Crystal structure discrimination scheme based on a single atom speed dynamics. a) Representative single atom speed trajectory from molecular dynamics simulations of bct5 (orange) and btin (blue) samples at phase transition point of 300 K and 16.5 GPa. b) Speed probability distribution function (PDF) determined on the basis of 20-ps-long speed trajectory of randomly selected bct5- and btin-atom with the corresponding theoretical Maxwell distribution (green). c) Distribution of Shannon entropies of speed's PDFs based on 20-ps-long trajectories of 56 randomly selected bct5- and btin-atoms. As the Maxwell distribution of speed is structure-independent, the distributions of the entropies overlap. d) Symbolic permutation-pattern-based representation of the single atom speed trajectory showed in figure a). e) Permutation patterns probability distribution function, so-called ordinal distribution, determined on the same trajectories as used in figure b). f) Distribution of the Shannon entropies of ordinal distributions based on the same trajectories as used in figure c). Non-overlapping distributions indicate different underlying crystal structures.}
\label{fig:scheme}
\end{sidewaysfigure}

\begin{figure}[h]
\centering
\includegraphics[width=\textwidth]{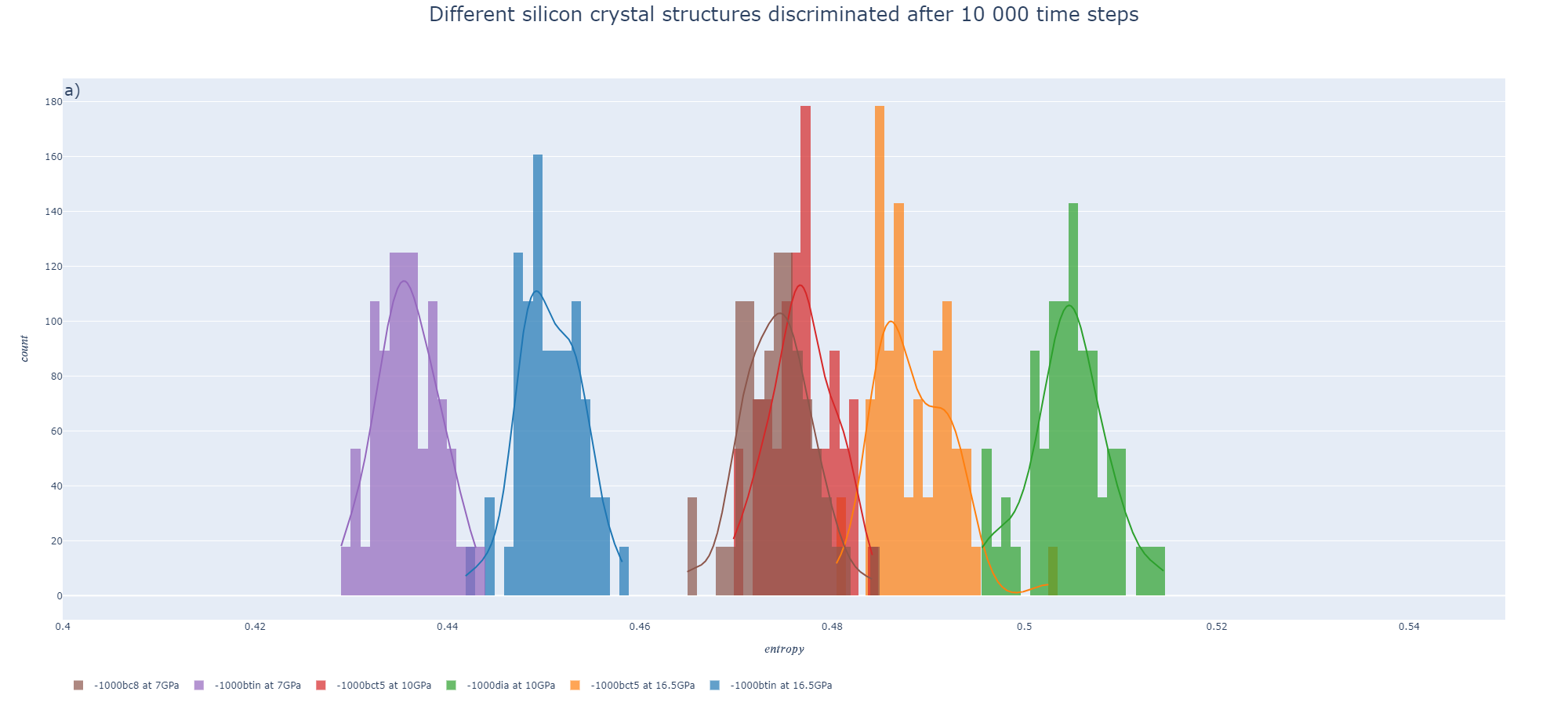}
	\caption{Discrimination of silicon crystals at different phase transition points. Permutation entropies distributions determined using the scheme presented in \hyperref[fig:scheme]{Figure 1} with 20-ps-long trajectories of 56 randomly selected atoms from every silicon samples simulated at 300K and different pressures: diamond and bct5 structures at 10 GPa, bct5 and btin at 16.5 GPa, btin and bc8 at 7GPa. The distributions that correspond to the phases modeled at the same pressure are clearly separated indicating the discriminative power of the method.}
\label{fig:all}
\end{figure}

Crystal structure discrimination scheme is presented in \hyperref[fig:scheme]{Figure 1}. Initially, we analyzed the speed trajectories of 56 randomly selected atoms from both bct5 and btin crystal samples simulated at 300 K and 16.5 GPa. Samples of single atom speed trajectories are shown in \hyperref[fig:scheme]{Figure 1a} with the corresponding distributions of speed presented in \hyperref[fig:scheme]{Figure 1b}. As the systems under study are ergodic, the probability of atom speed is described by Maxwell distribution regardless of a phase atom belongs to. Therefore, the entropies that we calculated for every single atom speed distributions are structure-independent as well, providing no discrimination power in detecting different underlying phases (\hyperref[fig:scheme]{Figure 1c}). It is worth noting that the distribution of speed is invariant with respect to the permutation of speed trajectory, indicating no internal temporal structure. However, the process of sampling speed from a given Maxwell distribution might depends on the underlying crystal phase. Here we quantified this process using permutation entropy, i.e. we generated symbolic speed trajectories (\hyperref[fig:scheme]{Figure 1d}) incorporating the temporal structures of atom speed dynamics, determined the probability distribution of permutation patterns (\hyperref[fig:scheme]{Figure 1e}) and finally calculated permutation entropy for every randomly selected atoms (\hyperref[fig:scheme]{Figure 1f}). In this case, compared to \hyperref[fig:scheme]{Figure 1c}, the distributions of permutation entropies clearly don't overlap indicating distinct underlying phases, as anticipated. Subsequently, we discriminated the other crystal structures using presented scheme as shown in \hyperref[fig:all]{Figure 2}.

\begin{figure}[h]
\centering
\includegraphics[width=\textwidth]{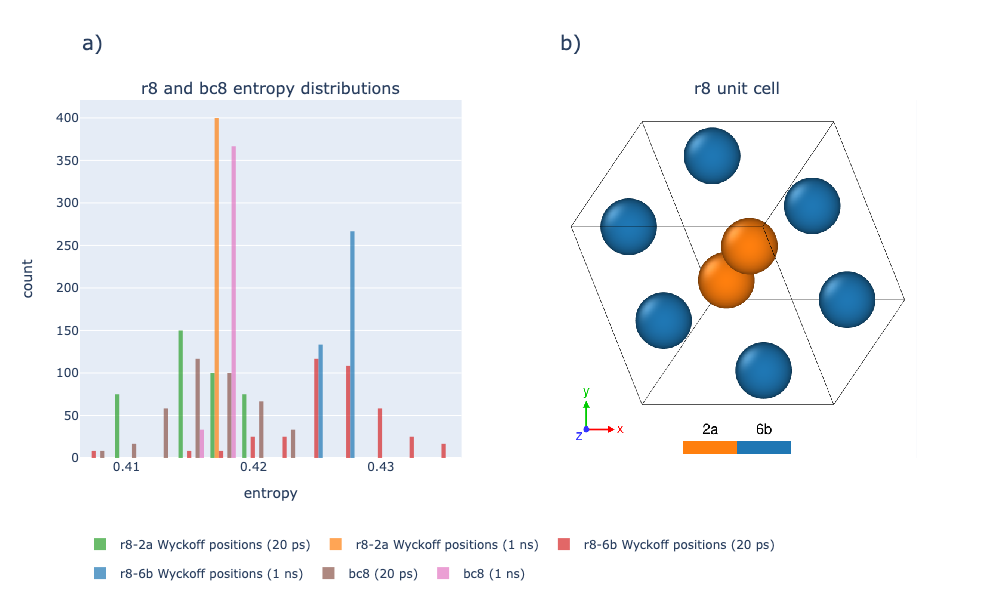}
	\caption{Discrimination of the Wyckoff positions of r8 and bc8 silicon crystal phases. a) Permutation entropies distributions determined using the scheme presented in \hyperref[fig:scheme]{Figure 1} with 20-ps- and 1-ns-long trajectories of 16 r8-2a-positions, 48 r8-6b-positions and 48 bc8-atoms from the piece of the simulated samples. A clear discriminations are achieved for 1 ns long trajectories and the overlapping of r8-2a-positions and bc8-atoms reflects the similarity of the phases. b) Unit cell of r8 crystal structure with 2a (orange) and 6b (blue) Wyckoff positions colored according to their permutation entropies.}
\label{fig:wyckoff}
\end{figure}

A clear separation of the permutation entropy distributions presented in \hyperref[fig:all]{Figure 2} was achieved using 10 000 steps long trajectories. In general, the longer the trajectory, the lower dispersion of distributions. It should be clear, however, that the trajectory's length must be much greater than the number of possible permutation patterns in order to obtain credible results. Moreover, the length of the trajectories required to discriminate subtle differences in local atomic arrangements might be even longer as in the case of r8 silicon structure that consists of two Wyckoff position, each of which with its unique neighborhood. This is demonstrated in \hyperref[fig:wyckoff]{Figure 3}. Similarly to the above analyzes, we initially used 10 000 long trajectories of both r8- and bc8-atoms, obtaining no clear discrimination between the phases. In turn, we analyzed 1 ns long trajectory of both structures that finally provided satisfactory results with two permutation entropy distributions for r8 structure \hyperref[fig:wyckoff]{Figure 3a} corresponding to two Wyckoff positions thereof. The crystal structure of r8 phase is shown in \hyperref[fig:wyckoff]{Figure 3b} with the Wyckoff positions indicated by colors. Additionally, it is worth noting that the permutation entropies of bc8-atoms overlap with one of the r8-distributions, reflecting subtle differences between these phases \cite{10.1103/PhysRevB.101.245203}.

\section{Discussion}\label{discussion}
The above results clearly indicate that a temporal structure of single atom speed trajectory can be effectively utilized to identify distinct local atom arrangements in defect-free crystals at equilibrium. Relaxing constrains on simulated samples to test the method against defected or disordered multi-element materials at different physical conditions (temperature, pressure, shear strain, etc.) is out of the scope of this work as well as using different simulation settings (eg. models or dynamics). As the method is based on single atom attributes, it would enable an efficient and precise identification of atom that initiates a~structural change within a material, enhancing our understanding of structure-function relationship. Further development of the method should be aimed at shortening trajectory length in order to efficiently study both materials phenomena out of equilibrium and non-ergodic systems. To this end one would use various symbolization techniques that quantify additional information (eg. amplitude-aware permutation patterns \cite{10.1103/PhysRevE.87.022911}), other atom trajectories (eg. positions, displacements, forces, etc.) or different statistical measures (eg. statistical complexity \cite{Lopez-Ruiz1995}, other entropies \cite{Pessa2021} ). Moreover, our results might be significantly improved by using advanced statistical or machine learning methods that utilize the temporal structure of trajectories, eg. ordinal networks \cite{Small2013, Bandt2023}, computational mechanics \cite{Crutchfield2012} or recurrent neural networks \cite{Muoz-Gil2021}. Finally, an especially promising approach to discriminate local atom arrangements in materials would incorporate both the temporal \textit{and} spatial structures of (multi-) atom trajectories.

\printbibliography

\end{document}